\documentclass[
    titlepage,
    superscriptaddress,
    twocolumn,
    a4paper,
    floatfix,
    longbibliography
]{revtex4-2}

\usepackage[american]{babel}
\usepackage[utf8x]{inputenc}

\usepackage{grffile} 

\usepackage{microtype}

\usepackage{amssymb}
\usepackage{bbm}
\usepackage{dsfont}
\usepackage{braket}
\usepackage{mathrsfs}

\usepackage[clock]{ifsym}

\usepackage{graphicx}
\usepackage[svgnames]{xcolor}

\usepackage{siunitx} 

\usepackage{ragged2e}
\usepackage{array}
\usepackage{tabularx}
\usepackage{booktabs}

\usepackage{siunitx}

\definecolor{cset-aps-blue}{RGB}{18,84,168}
\definecolor{cset-aps-limegreen}{RGB}{153,204,51}

\definecolor{cset-aps-blueberry}{RGB}{28,128,158}
\definecolor{cset-aps-turquoise}{RGB}{0,67,88}
\definecolor{cset-aps-limegreen}{RGB}{190,219,67}
\definecolor{cset-aps-darkblue}{RGB}{31,138,112}
\definecolor{cset-aps-yellow}{RGB}{255,225,25}
\definecolor{cset-aps-orange}{RGB}{253,116,0}
\definecolor{cset-aps-red}{RGB}{219,0,43}

\newcommand{\eg}[0]{\textit{e.\,g.}}
\newcommand{\ie}[0]{\textit{i.\,e.}}

\usepackage{hyperref}
\hypersetup{%
    colorlinks=true,
    linkcolor={cset-aps-red},
    linkbordercolor={cset-aps-red},
    filecolor={cset-aps-orange},
    filebordercolor={cset-aps-orange},
    citecolor={cset-aps-blue},
    citebordercolor={cset-aps-blue},
    urlcolor={cset-aps-blue},
    urlbordercolor={cset-aps-blue},
    menucolor={cset-aps-limegreen},
    menubordercolor={cset-aps-limegreen},
    breaklinks=true,
    pdfborderstyle={/S/U/W 2},
    pdfpagemode=UseOutlines,
    pdfstartpage={1},
}

\newcommand{\orcid}[1]{\href{https://orcid.org/#1}{\includegraphics[width=7pt]{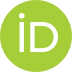}}}

\newcommand{\affTUDa}{\address{Technische Universit{\"a}t Darmstadt, Fachbereich Physik, Institut f{\"u}r Angewandte Physik, Schlossgartenstr. 7, 64289 Darmstadt, Germany}}

\begin{document}

\title[Baseline optimization]{Spatial and Pulse Efficiency Constraints in Atom Interferometric Gravitational Wave Detectors
}

\author{P~Schach\,\orcid{0000-0002-6672-9692} and E~Giese \orcid{0000-0002-1126-6352}}
\affTUDa

\collaboration{This article has been published in \href{https://doi.org/10.1088/2058-9565/adffb1}{Quantum Science and Technology \textbf{10}, 045031 [2025]} under the terms of the \href{https://creativecommons.org/licenses/by/4.0/}{Creative Commons Attribution License 4.0 [CC BY]}.}

\begin{abstract}
Currently planned and constructed terrestrial detectors for gravitational waves and dark matter based on differential light-pulse atom interferometry are designed around three primary strategies to enhance their sensitivity: 
(i) Resonant-mode enhancement using multiple diamonds, 
(ii) large-momentum-transfer techniques to increase arm separation within the interferometer, 
and (iii) very-long baseline schemes that increase the distance between the two interferometers.
Both resonant-mode enhancement and large-momentum-transfer techniques result in a greater number of light pulses, making high pulse fidelity during atom-light interactions imperative.
At the same time, increasing the number of diamonds in vertical configurations leads to taller atomic fountains, which consequently reduces the available distance between interferometers.
As a result, the number of diamonds, large-momentum-transfer pulses, and the fountain height are interdependent parameters that must be carefully balanced.
In this work, we present optimal configurations for multi-diamond geometries in resonant mode, explicitly accounting for the spatial extent of a single interferometer, considering constraints imposed by the baseline dimensions and atomic losses due to imperfect pulses.
For this optimization, we numerically scan the relevant parameters such as initial position and momentum of the atomic cloud, transferred momenta, and number of loops.  
For each parameter set, we verify whether the imposed conditions are met and evaluate the resulting sensitivities to identify optimal configurations.
We provide practical analytical relations to estimate the optimal number of pulses that should be applied.
Many proposals beyond demonstrator experiments require pulse numbers that demand efficiencies not yet demonstrated with state-of-the-art momentum transfer techniques. 
As a result, the observed sensitivity falls short of expectations---an effect caused by both arm separation and atom loss per pulse---highlighting the urgent need for research aimed at improving pulse fidelities.

\end{abstract}

\maketitle

\section{Introduction}
Terrestrial very-long baseline atom interferometers~\cite{abend2024, abdalla2025, balaz2025} for the detection of gravitational waves~\cite{dimopoulos2008, hogan_atomic_2011, graham2013, graham2016} and dark-matter candidates \cite{arvanitaki2018, badurina2022, derr2023} are currently planned or under construction, \eg{}, the MAGIS~\cite{abe2021}, AION~\cite{badurina2020}, ELGAR~\cite{canuel2020}, ZAIGA~\cite{zhan2020}, and MIGA~\cite{canuel2018} initiatives.
These instruments are designed to complement traditional optical interferometers by filling the sensitivity gap in the frequency range from \qty{0.01}{Hz} to a few Hertz. 
To suppress common-mode noise, the proposals employ differential measurements between two spatially-separated interferometers, which are aligned either vertically or horizontally. 
Other noise sources such as Newtonian or gravity-gradient noise~\cite{harms2015, carlton2023, carlton2025, saulson1984, canuel2020_arxiv} pose significant challenges for gravitational-wave detection in the low-frequency band of the targeted regime.
However, correlation methods~\cite{chaibi2016, badurina2023} offer an approach to reduce the impact of these noise sources.
Differential laser phase noise~\cite{gouet2007, bohringer2025} can be effectively suppressed by employing single-photon transitions~\cite{hu2017, hu2019, rudolph2020, bott2023}, while other setups relying on cavity-based two-photon transitions~\cite{canuel2018} must develop special techniques to address this challenge.

In principle, three primary strategies can be employed to enhance the sensitivity of the detectors.
The first involves increasing the number of atom–light interaction points by implementing large-momentum-transfer (LMT) methods.
The most prominent LMT techniques include Bloch oscillations~\cite{clade2009, mcdonald2013, pagel2020, rahman2024, fitzek2024}, double diffraction~\cite{leveque2009, giese2013, kuber2016}, higher-order diffraction~\cite{hartmann2020, muller2008}, and the application of sequential pulses~\cite{rudolph2020, berg2015, mcguirk2000}, which  are most commonly employed with single-photon transitions and which will be the focus of our discussion below.
The second method relies on resonant-mode amplification~\cite{graham2016} and is implemented by matching the frequency of the gravitational wave to the interrogation time of a single loop of a multi-diamond configuration~\cite{clauser1988, marzlin1996, kleinert2015, di_pumpo2023, schubert2021}.
As a result, the signal amplitude scales with the number of diamonds.
While in some configurations the roles of the interferometer arms are interchanged with each diamond, the majority of proposed schemes introduce a mirror pulse after each diamond, thereby preventing arm crossings and allowing the phase difference to accumulate coherently.
This latter scheme is the focus of the present article.
The third strategy improves sensitivity by using very large baselines, as this increases the difference between two local interferometers that interact with the gravitational wave at different points in space and time.
However, this approach is limited by spatial constraints, particularly for terrestrial and vertical configurations, where the interferometer arms may eventually reach the top or bottom of the available baseline.
In contrast, for horizontal configurations, neither the interferometer duration nor the fountain height directly reduces the available baseline length.
However, these parameters cannot be increased indefinitely as they are limited by the transverse dimensions of the detector and may have to resort to relaunch strategies~\cite{abend2016, schubert2024}.
All of these strategies are interdependent and must be carefully balanced to optimize overall detector performance, as we discuss in this article.

The total number of applied diffraction pulses is determined by the number of pulses per LMT sequence and diamonds, which in turn is related to the number of applied beam splitter or mirror sequences. 
As a general rule, a larger total number of pulses is typically desired.
However, the total number of pulses is significantly constrained by the pulse fidelity, as imperfections in each pulse reduce the number of detected atoms, ultimately degrading the sensitivity due to shot noise.
Hence, there must be a trade-off between signal enhancement achieved by increasing the number of LMT pulses or diamonds and the increase in shot noise caused by atom loss.
Previous work~\cite{di_pumpo2024} optimizes the fountain height for a given baseline, a fixed number of diamonds, and a specific number of LMT pulses.
In these studies, the number of diamonds and the number of LMT pulses were treated as independent, predetermined parameters, and the difference in arm lengths within the interferometer was not explicitly considered.

Imperfect fidelities limit the total number of pulses that can be distributed between diamonds and LMT pulses.
Simultaneously, operating a large number of diamonds in resonant mode increases the required fountain height, while employing a greater number of LMT pulses leads to larger arm separations within the interferometer.
Both effects result in the interferometer occupying greater fractions of the available baseline.
In contrast to terrestrial atom interferometers, these spatial constraints do not necessarily apply to space-based~\cite{el-neaj2020} or horizontal proposals~\cite{canuel2018, canuel2020}.
For example, in the MAGIS-space proposal~\cite{graham2016}, optimization parameters are constrained by the maximal number of pulses and the maximal interferometer duration, but there are no spatial restrictions imposed by the baseline, as the atom interferometers are placed on separate satellites.

In this article, we determine optimal configurations for differential geometries in resonant-mode operation based on experimentally available parameters. 
Previous optimizations have not explicitly accounted for the effects of arm separation within a single interferometer or imperfect pulse efficiency.
We explicitly incorporate spatial limitations introduced by the baseline for the fountain height and arm separation, as well as losses per pulse in the optimization process.
In contrast to previous approaches~\cite{di_pumpo2024}, we assume that the atom flux of the source is independent of the interferometer duration, implying an interleaved operation~\cite{savoie2018} if necessary.

After defining the considered configuration in section~\ref{sec:setup}, we optimize for the fountain height and total number of pulses for which the explicit derivation is outlined in appendix \ref{sec:appendix}.
Other quantities such as the number of diamonds and number of LMT pulses are derived from these optimized parameters.
In section~\ref{sec:anal_opt}, we establish concise analytical relations to estimate the optimal number of pulses within an experiment as a function of key experimental parameters.
To leading order, we observe that the optimal pulse number is determined solely by atomic losses per pulse.
Moreover, we derive in section~\ref{sec:role_arm_sep} analytical conditions to identify the parameter space where the arm separation becomes relevant.
Finally, in section~\ref{sec:num_opt} we incorporate the spatial constraints into a numerical optimization and compare the results to the analytical approach that neglects restrictions.  
We find that most proposals beyond demonstration experiments require pulse numbers that imply efficiencies that have not yet been demonstrated with state-of-the-art LMT techniques, highlighting the urgent need for research aimed at enhancing pulse fidelities. 
We summarize our results and establish their connection to previous studies in section~\ref{sec:conclussion}.

\section{Setup}
\label{sec:setup}

\begin{figure*}[ht]
	\begin{center}
		\includegraphics[width=1.0\textwidth]{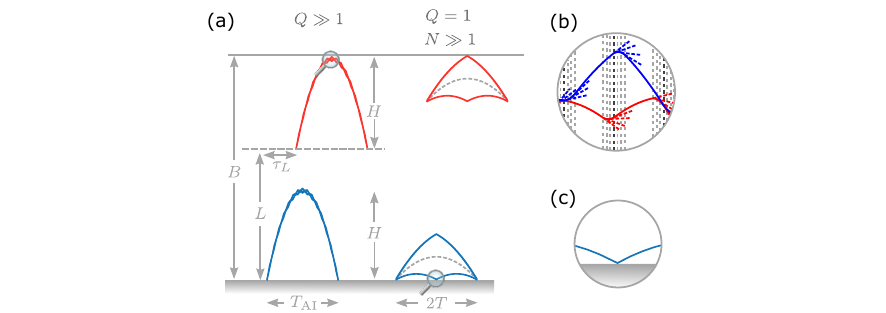}
		\caption{
		(a) Differential configuration of two atomic Mach-Zehnder interferometers, each with a fountain height $H$, separated by a distance $L$ and confined within a baseline of length $B$.
        Each interferometer consists of $Q$ diamonds, leading to a total interferometer time $T_{\text{AI}} = 2 Q T$ for an interrogation time $T$.
        The LMT beam splitters consist of $N$ single-photon pulses, whereby the mirrors are implemented via a sequence of $2N - 1$ single-photon pulses, see inset (b).
        The finite speed of light introduces a time delay $\tau_L$ between the lower and upper interferometer.
        The left panel illustrates the case of many diamonds, while the right panel depicts a single diamond.        
        (b) Visualization of the parasitic paths and atom losses that emerge from imperfect LMT pulses within a single diamond, showing the region magnified by the magnifying glass on the left of panel (a).
        The vertical extent of both interferometers is ultimately constrained by the bottom and top of the available baseline, so the atoms are ideally reflected before they hit the ground (c), which is the region magnified by the magnifying glass on the right of panel (a).
        } 
		\label{fig:sketch_diff_setup}
	\end{center}
\end{figure*}

In this article, we consider a differential configuration to measure phase fluctuations induced by a gravitational wave between two vertically-aligned Mach-Zehnder interferometers. 
In such a differential geometry, common-mode phase-noise contributions are reduced. 
To suppress remaining laser phase fluctuations, the beam splitter and mirror pulses in each interferometer are realized by single-photon transitions~\cite{hu2017, hu2019, rudolph2020, bott2023} with laser wave vector $k$ that corresponds to an optical wavelength.
The two interferometers are separated by a distance $L$ and confined within the baseline $B$, which represents the ultimate spatial resource in such very-large baseline setups, see figure~~\ref{fig:sketch_diff_setup}.
In addition, the baseline limits the other relevant spatial quantities, namely the maximum allowable arm separation and spatial extent or fountain height of the interferometers, fundamentally constraining the interferometer design and sensitivity.
For large baselines, the finite speed of light $c$ introduces significant time delays $\tau_L = L / c$ between both interferometers and $\tau_B = B / c$ along the whole baseline.
Effects due to finite speed of light over the extent of a single interferometer with fountain height $H = B - L$ are neglected in the following. 

Utilizing the full capabilities of differential Mach-Zehnder interferometers, two modes of operation are possible: 
(i) in resonant mode~\cite{graham2016} the interferometer time $T$ is matched to the frequency of a gravitational wave $f = \omega / (2 \pi)$ through the condition $\omega T = \pi$, 
(ii) while in broadband mode~\cite{badurina2023_broad} this condition is relaxed, allowing the interferometer to respond over a wider frequency range at the cost of reduced peak sensitivity.
To further amplify the gravitational wave signal, we allow for a multi-diamond scheme characterized by the number of diamonds $Q$.
This configuration increases the signal amplitude, effectively providing a $Q$-fold resonant enhancement.
However, this increased sensitivity comes with narrower resonances, reducing the bandwidth of the detector~\cite{graham2016} by $1/Q$ and making it more selective to specific frequencies.

Instead of applying a single light pulse for each beam-splitter and mirror, LMT techniques can be incorporated to significantly increase the spatial separation between the two arms within each diamond. 
Specifically, each beam splitter consist of $N$ single-photon pulses, whereby the mirrors are implemented via a sequence of $2N - 1$ single-photon pulses, resulting in an increase in the transferred momentum by the factor $N$.
To avoid spontaneous emission or noise arising from populating different internal states, we assume a even number of LMT pulses such that between two pulse sequences both interferometer arms are in the ground state.
Figure~\ref{fig:sketch_diff_setup}(a) visualizes the differential multi-diamond scheme for many diamonds (left) and a single diamond (right).

Generalizing the results for gravitational waves~\cite{graham2016}  in analogy to calculations for the sensitivity on dark matter~\cite{badurina2022}, the signal amplitude of the differential phase between both interferometers for a gravitational wave of strain $h$ in the low-frequency band for $\omega \tau_B \ll 1$ is given by
\begin{equation}
    \Phi = \frac{4 h k c}{\omega} \frac{L}{B} \left| 
    \sin \frac{\omega \tau_B N}{2}  \sin \frac{\omega T}{2}  
    \sin \frac{\omega \left[ T - (N - 1) \tau_B \right]}{2} 
    ~\frac{\sin\left( Q \omega T \right)}
    {\sin\left( \omega T \right) }
    \right| 
    \label{eq:signal_ampl_sinc_broad}
\end{equation}
and is obtained by averaging the second moment of the differential phase over uniformly distributed phases of the gravitational wave~\cite{arvanitaki2018}, which can be used to describe broadband mode operation.
In resonant mode~\cite{graham2016} with $\omega T = \pi$ and expanding $\omega \tau_B \ll 1$ but still keeping  $N \omega \tau_B $, the signal amplitude takes with $\sin (Q\pi)/\sin \pi =Q$ the form
\begin{equation}
    \Phi = \frac{4 h k c}{\omega} \frac{L}{B} Q \left| 
    \sin \frac{\omega \tau_B N}{2}  
    \cos \frac{\omega \tau_B N}{2}  
    \right| = 2 h k L N Q \left| \mathrm{sinc}\left( \omega \tau_B N \right) \right|,
    \label{eq:signal_ampl_sinc}
\end{equation}
where we used the trigonometric identity $2 \sin (N x/2) \cos (N x/2) / x =  \sin(N x )/x = N \mathrm{sinc}(Nx )$ in the last step.
For typical frequencies around~\cite{badurina2021} \qty{1}{Hz} and a baseline of~\cite{abe2021} \qty{100}{m}, the low-frequency approximation is well satisfied $\omega \tau_B\sim 10^{-6} \ll 1$. 
Assuming additionally a moderate number of pulses such that the condition $\omega \tau_B N \ll 1$ holds, the signal amplitude simplifies to
\begin{equation}
    \Phi \approx 2 h k L N Q .
\end{equation}
The corresponding strain sensitivity parametrized by the uncertainty 
\begin{equation}
    \Delta h = \frac{\Delta \Phi}{2 k L N Q} 
    \label{eq:sens_fun_approx}
\end{equation}
is readily derived from Gaussian error propagation.
Here, $\Delta \Phi$ denotes the uncertainty of the signal amplitude.

The signal amplitude is obtained from the phase difference measured between the two interferometers. 
This constitutes a correlation measurement~\cite{barrett2015} that inherently suppresses common-mode noise, a crucial method for achieving the sensitivities required for gravitational wave detection. 
Besides directly fitting the fringes, techniques such as ellipse fitting~\cite{foster2002} or Bayesian estimation~\cite{stockton2007} are routinely employed to infer the differential phase.
However, the precision of such correlation measurements is ultimately limited by shot noise, which acts independently in each atom interferometer.
We therefore assume that the uncertainty $\Delta \Phi$ of the signal amplitude is shot-noise limited or at least scales like shot noise with the number of atoms. 
Moreover, we consider that phase uncertainties arising from both external factors and technical noise sources can be efficiently mitigated.
The same reasoning is applied to Newtonian or gravity-gradient noise which limits sensitivity in the low-frequency band~\cite{beker2012, harms2015, carlton2025}, for example by relying on mitigation strategies like correlation methods~\cite{chaibi2016}.
Under these assumptions, the phase uncertainty becomes $\Delta \Phi = \sqrt{2 / (\nu N_\text{at} C^2})$ with the number of detected atoms $N_\text{at}$, number of repetitions $\nu$, and contrast $C$.
Here, the combination $\nu N_\text{at} = \dot{N}_\text{at} T_\text{int}$ can also be interpreted as the product of the atom source flux and the integration time $T_\text{int}$ of the measurement campaign.
In this latter form, effects such as possible dead times or an interleaved operation~\cite{savoie2018} of multiple interferometers simultaneously are, in principle, accounted for through averaged quantities.
This means that the effective repetition rate can be smaller than the total duration of a single interferometer cycle.
We observe that the strain uncertainty in equation~(\ref{eq:sens_fun_approx}) scales with the number of LMT pulses, which in turn contributes to the total number of applied light pulses $N_P$.
In experimental implementations, imperfect pulse efficiency induces atom loss, reducing the number of detected atoms according to $N_\text{at} = N_0 (1 - \lambda)^{N_P}$, where $\lambda$ is the loss per pulse.
Simultaneously, pulse imperfections may generate parasitic paths, as sketched in figure~\ref{fig:sketch_diff_setup}(b).
These unwanted paths can couple into the interferometer output port and fundamentally modify the interferometric response and contrast, an effect that we neglect here but that underlines the importance of mitigating such issues~\cite{pfeiffer2025, wilkason2022, rodzinka2024, wang2024, beguin2023}.
While the contrast will be influenced by these effects, particularly by the number of applied pulses or, more generally, by the interrogation time of the interferometer, we assume that these dependencies are sufficiently weak for the scope of this analysis. 
Nevertheless, we emphasize that they can be readily incorporated into the following treatment.
Incorporating atom losses, the strain uncertainty takes the form
\begin{equation}
   \Delta h = \frac{ \Delta \Phi (\lambda)}{2kLNQ} = \sqrt{\frac{2}{C^2\nu N_0 (1 - \lambda)^{N_P}}} \frac{1}{2kLNQ}
    \label{eq:dh_losses}
\end{equation}
with the initial number of atoms $N_0$ and relative losses per pulse $\lambda$.
In the following, we restrict our discussion to atom loss and exclude other mechanisms that may lead to decoherence or a reduction in contrast, specifically omitting the effects of parasitic path coupling into the exit port~\cite{beguin2022, altin2013, parker2016, jenewein2022}.
Here we observe a trade-off between two effects:
(i) atom losses increase the phase uncertainty by reducing the number of detected atoms, 
while (ii) a greater number of pulses per LMT sequence or more diamonds amplify the signal.
Previous optimization approaches did not account for this trade-off and typically assumed a fixed maximal number of pulses~\cite{di_pumpo2024}.
In contrast, we derive conditions incorporating both effects and identify optimal configurations.

\section{Analytical Optimization}
\label{sec:anal_opt}
Based on the strain uncertainty from equation~(\ref{eq:dh_losses}), we are able to identify optimal configurations for a given pulse efficiency. 
To obtain analytical relations, we make two assumptions, which we will relax later.
First, we connect the total interferometer duration $T_\text{AI} = 2 Q T$ to the fountain time $T_\text{tot} = \sqrt{8 H / g}$ of a fountain in a vertical setup.
Similarly to previous studies~\cite{di_pumpo2024}, we assume $T_\text{AI} = T_\text{tot}$, which leads to $Q = \xi \sqrt{\ell}$ with a scaling factor $\xi = \sqrt{2 B / (g T^2)}$ and the relative height $\ell = H / B$.
Because interleaved operation~\cite{savoie2018} is possible, we assume a fixed number of repetitions, so that the number of experimental repetitions $\nu$ is independent of the total interferometer time $T_\text{AI}$ and the interrogation time $T$.
The total number of pulses $N_P$ is a key parameter for the sensitivity and is related to the number of diamonds and LMT pulses by the relation $N_P = 4 Q N - 2 Q + 1$.
This assumption differs from previous treatments~\cite{di_pumpo2024}, in which $Q$ and $N$ were treated as independent quantities and the repetition rate was determined by the interferometer duration.
In addition, we assume lower bounds for the number of diamonds $Q=1$ and LMT pulses $N=2$, since the latter one has to be even.
Taking into account both assumptions, we optimize the sensitivity from equation~(\ref{eq:dh_losses}) with respect to the relative height $\ell$ and the total number of pulses $N_P$. 
The analytical relations for the optimal number of pulses $N_P$ and relative height $\ell$ are provided in \ref{sec:app_anal_losses}.
Based on these relations, the quantities $Q$ and $N$ are not necessarily integers.
Although the non-integer character of the number of diamonds and LMT pulses, which in the experiment is refined to even values, is not physical, the analytical relations for the optimal parameters nonetheless provide valuable insights into the optimization process.
In section~\ref{sec:num_opt}, we perform a numerical optimization with $Q$ and $N$ restricted to integer values and compare the results to the analytical observations.

\begin{figure*}[ht]
	\begin{center}
		\includegraphics[width=1.0\textwidth]{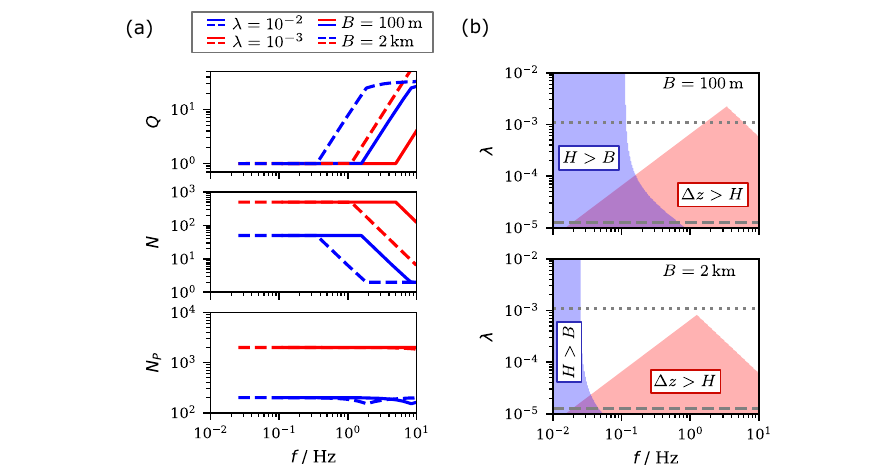}
		\caption{
		(a) Distribution of the optimal number of pulses $N_P$ between the number of diamonds $Q$ and LMT pulses $N$, as a function of the gravitational wave frequency $f$, shown for two different losses per pulse $\lambda$ (red and blue) and baselines $B$ (solid and dashed).   
        Assuming the entire fountain height can be exploited in an experiment and the arm separation can be neglected, we optimized the sensitivity from equation~(\ref{eq:dh_losses}) with respect to the relative fountain height $\ell=H/B$ and the total number of pulses $N_P$.
        (b) Taking into account the arm separation $\Delta z$ within a single interferometer, we find parameter spaces where the atomic cloud reaches the bottom (red shaded region) or top (blue shaded region) of the available baseline, and a numerical treatment becomes necessary.   
        } 
		\label{fig:anal_plots}
	\end{center}
\end{figure*}

The relation for the optimal number of pulses and its distribution between the number of diamonds and pulses per LMT sequence are plotted in figure~\ref{fig:anal_plots}(a).
We observe that a lower pulse efficiency (increased loss $\lambda$) leads to a reduced total number of pulses, which for low frequencies appears to be independent of both the baseline and the frequency.
Hence, in the low-frequency band of the targeted regime, the optimal number of pulses is primarily limited by the pulse efficiency. 

For small losses per pulse $\lambda \ll 1$, the total number of pulses in equation~(\ref{eq:anal_case2_np}) is approximated by 
\begin{equation}
    N_P \approx \frac{2}{\lambda} + \left( -\frac{1}{6} - \xi^2 \right) \lambda .
    \label{eq:approx_np}
\end{equation}
As expected, the pulse number is limited by the relative losses per pulse.
In particular, the approximation provides an estimate of the optimal number of pulses for given pulse efficiencies.
Conversely, for a specified number of pulses, the required pulse efficiency can be estimated to optimally utilize the available pulses.
The approximation consists of two terms: (i) the first, dominant term is independent of both baseline and frequency and inversely proportional to $\lambda$, while (ii) the second term scales linearly with $\lambda$, depends on baseline and frequency, and can become significant under certain conditions.

The total number of pulses, set by the pulse efficiency, can be allocated between increasing the number of LMT pulses (which results in greater arm separation) and the number of diamonds.
Since the sensitivity scales with the distance $L / B = (1 - \ell)$ between both interferometers and the number of diamonds $Q$, small fountain heights and large numbers of diamonds are desired.
However, the fountain height is connected to the number of diamonds by $Q \sim f \sqrt{\ell}$, leading to a trade-off between fountain height and number of diamonds.
Nevertheless, higher frequencies lead to shorter interrogation times $T = 1 / (2 f)$ in resonant mode, and consequently result in a greater number of diamonds for the same interferometer height.
Hence, we observe an increased number of diamonds for larger frequencies in the analytical optimization, visualized in figure~\ref{fig:anal_plots}(a).
While at low frequencies the number of diamonds is limited by a single diamond, at high frequencies there exists an upper bound for the optimal number of diamonds due to the constraint $N\geq 2$ and the relationship between the number of diamonds and LMT pulses through the total number of pulses.
The optimal pulse number is primarily determined by the pulse efficiency.

\section{Role of Arm Separation}
\label{sec:role_arm_sep}

So far, we have neglected the arm separation for each interferometer, which becomes increasingly relevant when considering the large number of LMT pulses proposed in recent schemes~\cite{badurina2020, abe2021}.
To check whether the trajectories implied by the analytical optimization for different pulse efficiencies and frequencies are confined within the baseline, we define the arm separation $\Delta z = N \hbar k T / m $ with the atomic mass $m$ and the maximal height of a fountain $H = \bar{\ell} B + \Delta z / 2$ where $\bar{\ell}$ is associated with the height of the mid-point trajectory.
Two scenarios can occur: the atom interferometer reaches (i) the ceiling or (ii) the bottom of the baseline.
To obtain an analytical condition for the first scenario, we assume low frequencies where a single diamond is optimal. 
Starting from the relation $Q = \xi \bar{\ell}^{1/2} = 1$, we find 
\begin{equation}
    \bar{\ell} = \frac{g}{8 B} \frac{1}{f^2} < 1 ,
    \label{eq:lower_bound_freq}
\end{equation}
which sets a lower limit on the frequency resolvable in resonant mode by a given baseline, visualized by the blue shaded regions in figure~\ref{fig:anal_plots}(b).
In the second scenario, the arm separation exceeds the fountain height $H / \Delta z > 1$, causing the lower interferometer arm to hit the bottom of the baseline, as visualized in figure~\ref{fig:sketch_diff_setup}(c). 
Using equations~(\ref{eq:anal_case2_np}), (\ref{eq:anal_case2_q}), and ~(\ref{eq:anal_case1_np}), we identify the parameter space where the interferometer is constrained by the bottom of the baseline.
In the limit $\lambda \ll 1$ and in the regime where a single diamond is optimal, the parameter space is restricted by $\lambda > f / \eta$ with $\eta =  g m / (\hbar k)$. 
In contrast, for $Q > 1$ and $N > 2$ the condition has the form
\begin{equation}
    \lambda > \left[ 
    \frac{8}{\eta} \left( \frac{g}{8 B} \right)^3 f^{-5} \right]^{\frac{1}{4}}.
    \label{eq:cond_param_hf}
\end{equation}
Both cases are visualized by the red shaded triangles in figure~\ref{fig:anal_plots}(b).

We observe that for the baseline $B=\qty{100}{m}$, the arm separation becomes relevant at lower pulse efficiencies and higher frequencies compared to the larger baseline $B=\qty{2}{km}$, consistent with the condition given in equation~(\ref{eq:cond_param_hf}).
Current proposals~\cite{abe2021} assume a large number of pulses per LMT sequence with $N = 4 \times 10^4$, resulting in a total number of pulses of approximately  $N_P \approx 1.6 \times 10^5$ for a single diamond.
Consulting equation~(\ref{eq:approx_np}) to optimally utilize the large number of pulses proposed, a loss of approximately $\lambda \approx 1.25 \times 10^{-5}$ per pulse is required.
The required efficiency is indicated by the dashed line in figure~\ref{fig:anal_plots}(b), which overlaps completely with the shaded regions corresponding to both baselines $B=\qty{100}{m}$ and $B=\qty{2}{km}$.
Consequently, the analytical expressions for the optimal parameters do not apply to the proposed number of pulses and their corresponding pulse efficiencies, given the spatial constraints imposed by the baseline.

\section{Numerical Optimization}
\label{sec:num_opt}

\begin{figure*}[ht]
	\begin{center}
		\includegraphics[width=1.0\textwidth]{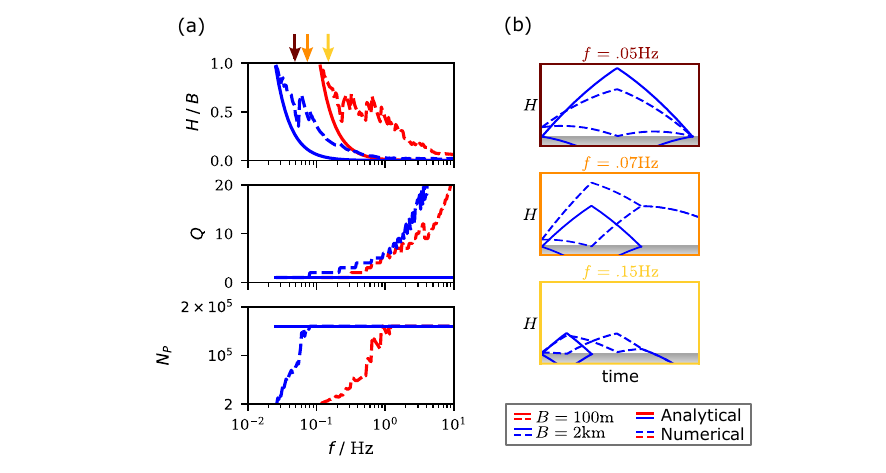}
		\caption{
        (a) Comparison of the optimal normalized fountain height $ H / B$, number of diamonds $Q$, and LMT pulses $N$, shown both with (dashed lines) and without (solid lines) accounting for arm separation as a function of the gravitational wave frequency $f$ for two different baselines $B$ (red and blue). 
        The analytical relations for the optimal parameters neglecting the arm separation are provided in appendix \ref{sec:appendix}.
        In contrast to the analytical optimization, the number of LMT pulses is constrained by the total number of pulses~\cite{abe2021} $N_P \approx 1.6 \times 10^5$, while the total interferometer time is limited by the fountain time.  
        (b) Visualization of optimal configurations for three different frequencies to illustrate the influence of arm separation on the optimization. 
        } 
		\label{fig:comp_anal_num}
	\end{center}
\end{figure*}
To incorporate the spatial restrictions imposed by the baseline, we transition from the analytical optimization to numerical methods. 
Relying on numerics, we loosen some of the assumptions made in the analytical optimization above.
Instead of treating the number of diamonds and LMT pulses as continuous parameters, we restrict $Q$ to positive integers and $N$ to positive even integers. 
Furthermore, the number of LMT pulses is not fixed but bounded from above by the total number of pulses according to $N < 0.5 + (N_P - 1) / (4 Q)$, where $N_P \approx 1.6 \times 10^5$ is derived from the proposed value $N = 4 \times 10^4$ for a single diamond.  
The same holds for the interferometer time, which does not have to equal the fountain time but is only limited by it from above.
To incorporate the timing of the first beam splitter into the optimization procedure, we introduce the initial height of each interferometer as an additional variable and optimize it jointly with the initial momentum of the atomic cloud, the number of diamonds, and the number of LMT pulses.
In contrast to the analytical optimization, the numerical optimization checks whether an atom interferometer reaches the lower or upper end of the baseline.  
Moreover, instead of considering the height of the midpoint trajectory as in the analytical optimization, we determine the actual height of a single atom interferometer and optimize the sensitivity from equation~(\ref{eq:signal_ampl_sinc}) without applying the approximation $\omega \tau_B N \ll 1$.

The relative fountain height, number of diamonds, and LMT pulses for optimal configurations depending on the frequency, both considering and neglecting arm separation, are compared in figure~\ref{fig:comp_anal_num} for both baselines $B=\qty{100}{m}$ and $B=\qty{2}{km}$.
Taking arm separation and baseline restrictions into account leads to larger relative heights and an increased number of diamonds with an increasing frequency.
In addition, peaks in the relative height arise because the optimal configuration no longer utilizes the entire fountain time.
As a consequence, the trajectory begins at larger heights and appears to bounce off the bottom of the baseline since the optimization simultaneously maximizes the number of LMT pulses while minimizing the relative height, as illustrated in figure~\ref{fig:comp_anal_num}(b). 

The effect of the top of the baseline can be observed in the number of LMT-pulses.
Up to a critical frequency, the number of LMT pulses remains constant.
However, below this frequency it decreases, because at low frequencies the upper arm of the interferometer reaches the top of the baseline.
To remain in resonant mode, the arm separation is decreased by reducing the number of LMT pulses.
This behavior is described by equation~(\ref{eq:lower_bound_freq}) and visualized by the blue shaded regions in figure~\ref{fig:anal_plots}(b).
There exists a minimal frequency below which the resonant-mode condition cannot be satisfied.
This critical frequency is higher for shorter baselines. 
To probe frequencies below this limit, it is necessary to switch to broadband mode.

In the regime of interest, we observe arm separations on the order of $H \approx 0.5 B$, which corresponds to a km-scale for a baseline of $B=\qty{2}{km}$.
This is several orders of magnitude larger than the arm separations achieved experimentally, which have reached up to half a meter~\cite{kovachy2015}.
In addition, the total interferometer time $T_{\text{AI}} = 2 Q T$ reaches approximately $\qty{5}{s}$ for a baseline of $B=\qty{100}{m}$ and exceeds $\qty{10}{s}$ for $B=\qty{2}{km}$.
To observe interference at the detectors, coherence times longer than $T_{\text{AI}} \approx \qty{10}{s}$ are necessary. 
Furthermore, to optimally utilize $N_P \approx 1.6 \times 10^5$, losses per pulse of $\lambda \approx 1.25 \times 10^{-5}$ are required, as estimated by equation~(\ref{eq:approx_np}).
This efficiency target is two orders of magnitude better than the current state-of-the-art losses per pulse of around~\cite{rodzinka2024} $\lambda = 1.1 \times 10^{-3}$.
Achieving these ambitious pulse efficiencies demands significant advances in pulse fidelity, coherence time, and collimation of the atomic cloud~\cite{ammann1997, muntinga2013, kovachy2015_lensing}.

\begin{figure*}[ht]
	\begin{center}
		\includegraphics[width=1.0\textwidth]{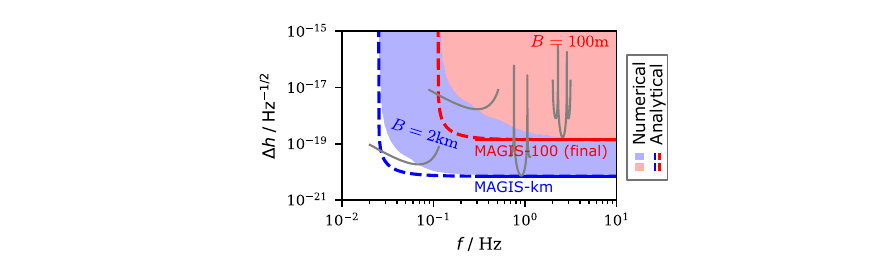}
		\caption{
		Comparison of the optimal achievable strain uncertainty $\Delta h$ for two different baselines as a function of gravitational wave frequency.
        The shaded regions represent the achievable sensitivity taking into account the arm separation.
        The dashed lines explicitly neglect it and therefore overestimate the achievable sensitivity.
        Optimal parameters are visualized in figure~\ref{fig:comp_anal_num}, and analytical relations neglecting the arm separation are provided in appendix \ref{sec:appendix}.
        The gray solid lines visualize the strain uncertainty for deviations from the resonant frequency, determined by Eq.~(\ref{eq:signal_ampl_sinc_broad}) using the optimal parameters corresponding to the frequencies \qty{0.07}{Hz} (with $Q=1$), \qty{0.30}{Hz} (with $Q=1$), \qty{0.91}{Hz} (with $Q=5$), and \qty{2.58}{Hz}  (with $Q=9$).
        In contrast to the analytical optimization, the number of LMT pulses is constrained by the total number of pulses~\cite{abe2021} $N_P \approx 1.6 \times 10^5$, while the total interferometer time is limited by the fountain time.  
        The solid lines represent the sensitivities targeted by the MAGIS project~\cite{abe2021} above \qty{0.3}{Hz}, rescaled to two atomic sources to match the optimization.
        } 
		\label{fig:sens_proposal}
	\end{center}
\end{figure*}

The optimal parameters are visualized in figure~\ref{fig:comp_anal_num}, and analytical relations for the case neglecting arm separation are provided in appendix \ref{sec:appendix}. 
In contrast to the analytical optimization, numerical methods are used to incorporate arm separation as well as upper bounds on the number of large-momentum-transfer (LMT) pulses and the total interferometer duration.

After analyzing the parameters of optimal configurations that respect the arm separations of a single interferometer, we now discuss the effects on the optimized strain uncertainty. 
To calculate the sensitivity, we take the phase uncertainty $\Delta \Phi = \qty{e-5}{Hz^{-1/2}}$ for both baselines, which already accounts for a possibly imperfect contrast without the need to specify dead times, repetition rates, or interleaved operation and is commonly assumed in recent proposals~\cite{abe2021}.
Figure~\ref{fig:sens_proposal} compares the sensitivities obtained with and without accounting for arm separation within the optimization. 
For comparison, the sensitivities assumed from the MAGIS proposal~\cite{abe2021} are rescaled to two atomic sources to match the optimization in this work, visualized as dotted lines in figure~\ref{fig:sens_proposal}. 
At high frequencies, our results recover the projected sensitivities, whereas at lower frequencies, significant deviations are observed.
In the low-frequency band of the targeted regime, the arm separation becomes a crucial limiting factor.
As the top of the baseline constrains the spatial extent of the interferometer arms, the number of LMT pulses must decrease, leading to a significant reduction in sensitivity.
Our optimization procedure optimizes the configuration for a specific frequency by operating in resonant mode, which is of particular interest for larger $Q$.  
However, deviations from this resonant frequency experience a rapidly decaying strain sensitivity, as the bandwidth of the resonance decreases inversely with the number of diamonds~\cite{graham2016}, scaling as $1/Q$.  
To highlight the bandwidth for representative resonant frequencies, we present the corresponding sensitivities for the optimized parameters beyond the resonant frequencies by employing Eq.~(\ref{eq:signal_ampl_sinc_broad}).  
These frequency bands are indicated in Fig.~\ref{fig:sens_proposal} by gray lines.  
For both baseline values, the left curves correspond to parameters with $Q=1$, while on the right we observe a narrowing of the resonance width for $Q=5$ (with baseline $B=2$\,km)  and $Q=9$ (with baseline $B=100$\,m).
Hence, the optimal sensitivity is limited to the frequency band around the resonance frequency for multiple diamonds, which confines the possible sources that can be detected.  
However, since the obtained values for $Q$ are still small as additional diamonds are costly in terms of LMT pulses, the decrease in bandwidth might be tolerable.
For $Q=1$ on the left flank of the sensitivity, where the height of the fountains becomes unphysical, we observe that broadband mode in fact outperforms the optimized resonant-mode scheme.  
We return to this point in the conclusions.

\begin{figure*}[ht]
	\begin{center}
		\includegraphics[width=1.0\textwidth]{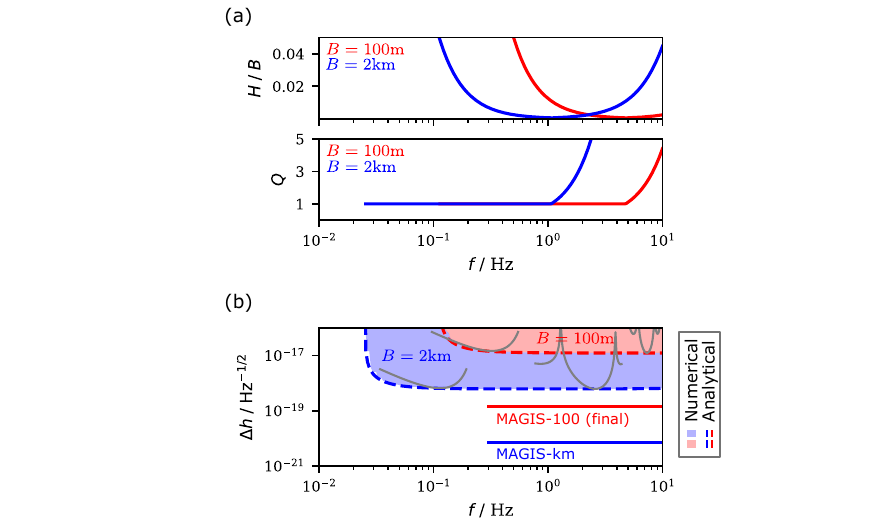}
		\caption{
        (a) Optimal normalized fountain height $H / B$ and number of diamonds $Q$, as well as (b) the corresponding sensitivity for two different baselines as a function of gravitational wave frequency.  
        The number of LMT pulses is constrained by a fixed total number of pulses, estimated using equation~(\ref{eq:approx_np}) for the losses $\lambda_R = 1.1 \times 10^{-3}$ corresponding to the current state of the art~\cite{rodzinka2024}.   
        Analytical relations for the case of a fixed total number of pulses are provided in \ref{sec:app_lossless}.
        The gray solid lines visualize the strain uncertainty in broadband mode, determined by Eq.~(\ref{eq:signal_ampl_sinc_broad}) using the optimal parameters corresponding to the resonant frequencies \qty{0.11}{Hz} (with $Q=1$), \qty{0.32}{Hz} (with $Q=1$), \qty{2.58}{Hz} (with $Q=2$), and \qty{7.32}{Hz}  (with $Q=6$).
        (b) The dashed lines represent the achievable sensitivity neglecting the arm separation, whereas the shaded regions explicitly take it into account.
        The solid lines indicate the sensitivities targeted by the MAGIS project~\cite{abe2021} above \qty{0.3}{Hz}, rescaled to two atomic sources to match the optimization.
        Even though our estimates are less sensitive than the target values, they remain several orders of magnitude below the value of $5.3\times10^{-15}\, \text{Hz}^{-1/2}$ assumed for the 100-m baseline, corresponding to a phase uncertainty of $\Delta \Phi= 10^{-3}\,\text{rad} \,\text{Hz}^{-1/2}$ and state-of-the-art LMT technology with 100 LMT pulses~\cite{abe2021}.
        } 
		\label{fig:sens_realistic}
	\end{center}
\end{figure*}

In the following, we assume the relative losses per pulse of $\lambda_R = 1.1 \times 10^{-3}$ corresponding to the current state of the art for Bragg interferometers~\cite{rodzinka2024}, which has been achieved using Floquet state engineering in optical lattices.  
While such a value has yet to be demonstrated in LMT sequences employing single-photon transitions, which currently~\cite{rudolph2020} exhibit per-pulse losses of $0.011$, one order of magnitude larger, we nonetheless take the result of this lattice sequence.
It provides an ambitious but realistic order of magnitude, even though this performance may not be directly applicable to single-photon transitions.
Using equation~(\ref{eq:approx_np}), the optimal number of pulses can be estimated, where the term linear in $\lambda$ scaling with $\xi^2 \propto B f^2$ becomes relevant. 
At frequencies $f < \qty{1}{Hz}$ and pulse-per-pulse atom loss $\lambda_R$, the optimal total number of pulses is approximately $N_P \approx 1800$ for both baselines $B=\qty{100}{m}$ and $B=\qty{2}{km}$.
However, for the larger baseline of \qty{2}{km} and higher frequencies, the linear term becomes crucial, reducing the optimal pulse number to about $N_P \approx 1640$ at frequencies around $f \approx \qty{10}{Hz}$.
Nevertheless, the total number of pulses is much smaller than assumed in the MAGIS proposal~\cite{abe2021} and used for the optimization above.
Consulting figure~\ref{fig:anal_plots}(b), the relative loss $\lambda_R$ indicated by the dotted gray line shows that the restrictions imposed by the baseline and arm separation are not significant in this regime.

The optimized sensitivities, both including and excluding arm separation, are compared in figure~\ref{fig:sens_realistic}.
The analytical and numerical optimizations show good agreement, indicating that the effect of arm separation is negligible in the considered parameter regime.
Due to the negligible effect of the arm separation, figure~\ref{fig:sens_realistic}(a) shows only the parameters obtained from the analytical formulas. 
In the frequency regime of interest, the relative height is approximately $5\,\%$ of the baseline, which is significantly smaller than the proposed parameters, as shown in figure~\ref{fig:comp_anal_num}. 
Consequently, the arm separation plays a negligible role in the optimization of the multi-diamond geometries.
Even though our estimates are less sensitive than targeted by the MAGIS proposal, they remain several orders orders of magnitude below the value of $5.3\times10^{-15}\, \text{Hz}^{-1/2}$ assumed for the 100-m baseline, corresponding to a phase uncertainty of $\Delta \Phi= 10^{-3}\,\text{rad} \,\text{Hz}^{-1/2}$ and state-of-the-art LMT technology with 100 LMT pulses~\cite{abe2021}. 

\section{Conclusion}
\label{sec:conclussion}
In this article, we have derived the optimal value for the total number of pulses in resonant mode, which depends on the experimental parameters and constraints.
It arises from a trade-off between maximizing the number of detected atoms, which is limited by the losses per pulse, and increasing the number of diamonds or LMT pulses to enhance the sensitivity.
We have derived analytical relations to estimate optimal pulse numbers for given experimental parameters such as losses per pulse and available baseline.
Notably, in the low-frequency band of the targeted regime, the optimal pulse number is primarily determined by atom loss.
Small losses allow for large pulse numbers, resulting in greater arm separations, which are fundamentally constrained by the baseline.
By explicitly incorporating the spatial limitation imposed by the baseline, we observed a significant reduction in sensitivity in the low-frequency band.
So far, neither the losses per pulse nor the arm separation of a single interferometer has been considered in the estimation of proposed sensitivities.

Moreover, pulse imperfections can generate parasitic paths, which reduce sensitivity by in-coupling into the exit port~\cite{beguin2022, altin2013, parker2016, jenewein2022}.
To mitigate these effects, several strategies can be employed, including the design of velocity-selective dichroic mirror pulses~\cite{pfeiffer2025}, tailored design of pulse shapes~\cite{saywell2020, wilkason2022, rodzinka2024}, optimal-control techniques~\cite{wang2024}, destructive interference of parasitic paths, and coherent enhancement~\cite{beguin2023}.
However, the contrast of the interferometer will depend on the number of applied pulses and experimental techniques, as well as the interrogation time and the temperature of the atomic cloud.
For a more precise optimization, a detailed model describing the dependence of the contrast on all these parameters can be incorporated, thereby extending our analysis.

Another limitation is imposed by the coherence time, expansion duration, temperature, and collimation of the atomic cloud.
In optimized configurations, interferometer times can reach $\qty{5}{s}$ for a baseline of $B\sim\qty{100}{m}$ and exceed $\qty{10}{s}$ for $B\sim\qty{2}{km}$, with corresponding arm separations on the order of \qty{}{km}. 
These values surpass current state-of-the-art capabilities in both arm separation~\cite{kovachy2015} and expansion duration of the atomic cloud by orders of magnitude.
To account for these constraints, it is necessary to impose upper limits on the total atom interferometer duration and arm separation.
In addition, this underlines the need for techniques like delta-kick collimation~\cite{ammann1997, muntinga2013, kovachy2015_lensing}.

When considering state-of-the-art losses per pulse and the corresponding optimal number of pulses, the arm separation plays a negligible role and does not currently present a limitation.
While our estimates do not reach the sensitivities targeted in the MAGIS proposal above \qty{0.3}{Hz} for a 100-m baseline, our results demonstrate that, with state-of-the-art LMT technology, $N_P=1800$ LMT pulses can be achieved---an order of magnitude improvement over $N_P=100$ considered in the initial stage of MAGIS~\cite{abe2021}.
This indicates that already current technology enables sensitivities better than anticipated by these conservative estimates.

An additional strategy to increase sensitivity is to enhance the signal utilizing entangled atoms~\cite{hosten2016, anders2021, cassens2025}, which enables sensitivities below the shot-noise limit.
We expect that the use of entanglement will not directly affect the optimization procedure.
However, losses per pulse can rapidly destroy the entanglement, thereby reducing its potential advantage~\cite{gunther2024}. 
The compatibility of entanglement-enhanced detection with LMT and multi-diamond schemes in long-baseline setups remains to be verified.

In this work, we have considered the effects of finite speed of light only on the scale of the baseline and between both interferometers. 
For a large number of LMT pulses, the finite speed of light also becomes relevant on the scale of a single interferometer, resulting in additional phase contributions~\cite{niehof2025}.
These contributions can be minimized by adjusting the resonant condition and by considering time-asymmetric configurations~\cite{tan2017}.

We emphasize that our optimization procedure was performed in resonant mode where $\omega T = \pi$, suitable only for a narrow frequency band around a known frequency of interest and thereby limiting the applicability of our results.  
In principle, one could extend our treatment by optimizing the interrogation time as an additional parameter.  
However, to optimize the sensitivity in broadband mode over a large frequency band, frequency-dependent optimization should be avoided, which requires different approaches and is therefore beyond the scope of the present article. 
However, even in our current treatment, we observe that for $Q=1$ there exists a cutoff frequency below which it is beneficial to resort to broadband mode, where the interrogation time is not linked to the frequency of interest.  
In fact, optimizing the baseline in this low-frequency regime requires additional analysis beyond the scope of the present work.

While our primary focus has been on gravitational-wave detectors, atom interferometers driven by single-photon transitions are also susceptible to ultralight scalar dark matter candidates~\cite{di_pumpo2022, derr2023}, exhibiting a sensitivity analogous to that for gravitational waves~\cite{dimopoulos2008, arvanitaki2018, badurina2021}
Therefore, our results can also be applied to the optimization~\cite{di_pumpo2024} of atom-interferometric dark-matter detectors.

\appendix
\section{Analytical optimization}
\label{sec:appendix}

\subsection{Case including atom loss}

\label{sec:app_anal_losses}

The strain uncertainty obtained by Gaussian error propagation is given by
\begin{equation}
    \Delta h = \sqrt{\frac{2}{C^2\nu N_0 (1 - \lambda)^{N_P}}} \frac{1}{2 k L N Q}
\end{equation}
with the initial number of atoms $N_0$, the relative losses per pulse $\lambda$, the interferometric contrast $C$, and the wave vector $k$ of the laser pulse.
The differential measurement scheme employs two interferometers separated by a distance $L$, with the height $H$ of each interferometer fountain contained within the baseline $B$.
The total number of pulses $N_P$ is related to the number of diamonds $Q$ and the number $N$ of LMT pulses per diffraction sequence through the relation $N_P = 4 Q N - 2 Q + 1$.

Furthermore, we impose two conditions: 
(i) The interferometer time is chosen as $T_\text{AI} = 2 Q T$, equal to the fountain time $T_\text{tot} = \sqrt{8 H / g}$.
This leads to the relation $Q = \xi \sqrt{\ell}$ with $\xi = \sqrt{2 B / (g T^2)}$ and $\ell = H / B$. 
(ii) We assume the maximal exploitation of the total number of pulses. 

Taking these assumptions into account, the strain uncertainty can be written as
\begin{equation}
    \Delta h \propto 
    \left[ 
    \left( 1 - \lambda \right)^{\frac{N_P - 1}{2}}
    \left( 1 - \ell \right)
    \left(\xi \sqrt{\ell} +  \frac{N_P - 1}{2}\right)
    \right]^{-1} . 
    \label{eq:anal_rel_to_opt}
\end{equation}
We optimize equation~(\ref{eq:anal_rel_to_opt}) with respect to the relative fountain height $\ell$ and the total number of pulses $N_P$. 
The minimal strain uncertainty is observed for an optimal height
\begin{equation}
    \sqrt{\ell} = 
    \frac{1}{\xi \mathrm{log}\left( 1 - \lambda \right)} +
    \sqrt{\left(\frac{1}{\xi \mathrm{log}\left( 1 - \lambda \right)} \right)^2 + 1}
\end{equation}
and an optimal total number of pulses
\begin{equation}
    N_P = 
    \frac{-4}{\mathrm{log}\left( 1 - \lambda \right)} -
    2 \sqrt{\left(\frac{1}{\mathrm{log}\left( 1 - \lambda \right)} \right)^2 + \xi^2} + 1,
    \label{eq:anal_case2_np}
\end{equation}
reducing to $N_P \approx \frac{2}{\lambda} + \left( -1/6 - \xi^2 \right) \lambda$ for small loss $\lambda \ll 1$.
Derived from the relative height, the optimal number of diamonds yields
\begin{equation}
    Q = \xi \sqrt{\ell} =
    \frac{1}{\mathrm{log}\left( 1 - \lambda \right)} +
    \sqrt{\left(\frac{1}{\mathrm{log}\left( 1 - \lambda \right)} \right)^2 + \frac{8 B}{g} f^2}
    \label{eq:anal_case2_q} .
\end{equation}

We observe that for higher frequencies, the number of diamonds increases. 
Since $Q$ and $N$ are related through the total number of pulses, namely $N_P = 4 Q N - 2 Q + 1$, the number of LMT pulses per diamond decreases.
However, at least $N=2$ is required, defining the high-frequency regime. 
In contrast, at lower frequencies, the number of diamonds decreases, resulting in a low-frequency regime where the optimal configuration is achieved with the minimal number of diamonds, namely $Q=1$.

In this low-frequency regime, \ie{} $Q=1$, the relative height is described by $\sqrt{\ell} = 1 / \xi$, and the total number of pulses takes the form
\begin{equation}
    N_P = \frac{-2}{\mathrm{log}\left( 1 - \lambda \right)} - 1 
    \approx
    \frac{2}{\lambda} .
    \label{eq:anal_case1_np}
\end{equation}
In the high-frequency regime, \ie{} $N = 2$, the total number of pulses can be expressed by $N_P - 1 = 6 \xi \sqrt{\ell}$, and the relative height is defined by the nonlinear equation
\begin{equation}
    \Delta h \propto 
    \left[ 
    4 \xi \sqrt{\ell} 
    \left( 1 - \lambda \right)^{3 \xi \sqrt{\ell}} 
    \left(1 - \ell \right)
    \right]^{-1} .
\end{equation}

\subsection{Lossless case}
\label{sec:app_lossless}

The lossless case $\lambda  = 0$ must be considered separately. 
Because the phase uncertainty no longer depends explicitly on $N_P$, there is no well-defined optimal number of pulses.
Therefore, we assume a fixed $N_P$ and optimize solely with respect to the relative height $\ell$.
The strain uncertainty in the lossless case then takes the form
\begin{equation}
    \Delta h \propto 
    \left[ 
    \left( 1 - \ell \right)
    \left(\xi \sqrt{\ell} +  \frac{N_P - 1}{2}\right)
    \right]^{-1}
\end{equation}
and the optimal relative height is given by 
\begin{equation}
    \sqrt{\ell} = 
    \sqrt{\frac{1}{3} + \left( \frac{N_P - 1}{6 \xi} \right)^2} - \frac{N_P - 1}{6 \xi} .
\end{equation}

\section*{Acknowledgements}
We thank O. Buchmüller, C. McCabe, F. Di Pumpo, A. Friedrich, and T. Kovachy for stimulating discussions and suggestions.
We acknowledge contributions from the Terrestrial Very-Long-Baseline Atom Interferometry (TVLBAI) proto-collaboration.
The QUANTUS-VI project is supported by the German Space Agency at the German Aerospace Center (Deutsche Raumfahrtagentur im Deutschen Zentrum f\"ur Luft- und Raumfahrt, DLR) with funds provided by the Federal Ministry for Economic Affairs and Climate Action (Bundesministerium für Wirtschaft und Klimaschutz, BMWK) due to an enactment of the German Bundestag under grant no. 50WM2450E (QUANTUS-VI).

\section*{Author contributions}
Conceptualization: PS and EG.
Formal analysis: PS.
Funding acquisition: EG.
Methodology: PS and EG.
Investigation: PS.
Validation: EG.
Visualization: PS.
Supervision: EG.
Writing—original draft: PS.
Writing—review \& editing: PS and EG.

\section*{Competing interests}
The authors declare that they have no competing interests.

\section*{Data and materials availability}
All data needed to evaluate the conclusions in the paper are present in the paper. 

\section*{References}
\bibliography{BaseOpt}     

\end{document}